\documentclass[sigconf]{acmart}

\usepackage{booktabs} 
\usepackage{array}
\usepackage{multirow}

\usepackage{enumitem}
\usepackage{colortbl}
\usepackage{wasysym}
\usepackage{xcolor}
\usepackage{url}
\usepackage{soul}

\AtBeginDocument{%
  \providecommand\BibTeX{{%
    \normalfont B\kern-0.5em{\scshape i\kern-0.25em b}\kern-0.8em\TeX}}}

\copyrightyear{2025}
\acmYear{2025}
\setcopyright{acmlicensed}\acmConference[CHI '25]{CHI Conference on Human Factors in Computing Systems}{April 26-May 1, 2025}{Yokohama, Japan}
\acmBooktitle{CHI Conference on Human Factors in Computing Systems (CHI '25), April 26-May 1, 2025, Yokohama, Japan}
\acmDOI{10.1145/3706598.3714210}
\acmISBN{979-8-4007-1394-1/25/04}

\def \debug{}

\def \debug{}

\ifx \debug \undefined

\newcommand{\fx}[1]{{}}
\newcommand{\fixmesb}[1]{{}}
\newcommand{\fixmejc}[1]{{}}
\newcommand{\fixmezz}[1]{{}}
\newcommand{\fixmejX}[1]{{}}
\newcommand{\fixmech}[1]{{}}
\newcommand{\fixmesl}[1]{{}}
\newcommand{\rev}[1]{{#1}}
\else

\newcommand{\fx}[1]{{\textcolor{red}{#1}}}
\newcommand{\fixmesb}[1]{{\bf\textcolor{red}{ [ sB FIXME: #1 ]}}}
\newcommand{\fixmejc}[1]{{\bf\textcolor{blue}{ [ jC FIXME: #1 ]}}}

\newcommand{\fixmech}[1]{{\bf\textcolor{purple}{ [ cH FIXME: #1 ]}}}
\newcommand{\fixmesl}[1]{{\bf\textcolor{yellow}{ [ sL FIXME: #1 ]}}}
\newcommand{\rev}[1]{{\color{black}{#1}}}

\newcommand{\bma}{Be~My~AI}
\newcommand{\sbma}{BMA}  

\fi
\newcolumntype{L}[1]{>{\raggedright\let\newline\\\arraybackslash\hspace{0pt}}m{#1}}
\newcolumntype{C}[1]{>{\centering\let\newline\\\arraybackslash\hspace{0pt}}m{#1}}
\newcolumntype{R}[1]{>{\raggedleft\let\newline\\\arraybackslash\hspace{0pt}}m{#1}}
\newcounter{challengeno}

\newcounter{scenariono}

\newcounter{ideano}

\begin{document}


\title[Insights from Smartphone Interaction of Visually Impaired Users with Large Multimodal Models]
{Beyond Visual Perception: Insights from Smartphone Interaction of Visually Impaired Users with Large Multimodal Models}



\author{Jingyi Xie}
\affiliation{%
  \institution{Pennsylvania State University}
  \city{University Park}
  \state{PA}
  \country{USA}
  \postcode{16801}
}
\email{jzx5099@psu.edu}

\author{Rui Yu}
\affiliation{%
 \institution{University of Louisville}
  \city{Louisville}
  \state{KY}
  \country{USA}
  \postcode{40292}
}
\email{rui.yu@louisville.edu}

\author{He Zhang}
\affiliation{%
 \institution{Pennsylvania State University}
  \city{University Park}
  \state{PA}
  \country{USA}
  \postcode{16801}
}
\email{hpz5211@psu.edu}

\author{Syed Masum Billah}
\affiliation{%
  \institution{Pennsylvania State University}
  \city{University Park}
  \state{PA}
  \country{USA}
  \postcode{16801}
}
\email{sbillah@psu.edu}

\author{Sooyeon Lee}
\affiliation{
    \institution{New Jersey Institute of Technology}
  \city{Newark}
  \state{NJ}
  \country{USA}
  \postcode{07102}
}
\email{sooyeon.lee@njit.edu}

\author{John M. Carroll}
\affiliation{%
 \institution{Pennsylvania State University}
  \city{University Park}
  \state{PA}
  \country{USA}
  \postcode{16801}
}
\email{jmc56@psu.edu}




\begin{abstract} %

Large multimodal models (LMMs) have enabled new AI-powered applications that help people with visual impairments (PVI) receive natural language descriptions of their surroundings through audible text. We investigated how this emerging paradigm of visual assistance transforms how PVI perform and manage their daily tasks. Moving beyond usability assessments, we examined both the capabilities and limitations of LMM-based tools in personal and social contexts, while exploring design implications for their future development. Through interviews with 14 visually impaired users of Be My AI (an LMM-based application) and analysis of its image descriptions from both study participants and social media platforms, we identified two key limitations. First, these systems' context awareness suffers from hallucinations and misinterpretations of social contexts, styles, and human identities. Second, their intent-oriented capabilities often fail to grasp and act on users' intentions. Based on these findings, we propose design strategies for improving both human-AI and AI-AI interactions, contributing to the development of more effective, interactive, and personalized assistive technologies.

\end{abstract}




\maketitle

\section{Introduction}

People with visual impairments (PVI) face challenges in perceiving their surroundings due to the absence of visual cues. \rev{Traditional} AI-powered \rev{systems like} Seeing AI~\cite{SeeingAI2020} \rev{and} human-assisted \rev{services like} remote sighted assistance~\cite{BeMyEyes2020} \rev{have helped PVI interpret their environment and complete} daily tasks. \rev{Recent advances} in computer vision (CV) and natural language processing (NLP) \rev{have enabled AI systems to identify} objects and text \rev{in scenes while responding to photo-based queries from PVI~\cite{ahmetovic2020recog, mukhiddinov2022automatic, hong2022blind, morrison2023understanding, gonzalez2024investigating}.
The emergence of} large multimodal models (LMMs)~\cite{yu2023mm}, \rev{particularly} GPT-4~\cite{achiam2023gpt}, \rev{has transformed} visual question-answering (VQA) \rev{capabilities.} Researchers have begun \rev{exploring how these powerful LMMs might benefit} PVI~\cite{zhao2024vialm, yang2024viassist}. \rev{At the forefront of this development is} \textit{\bma{}} (\sbma{})~\cite{bma_usecase}, the first \rev{publicly-available} LMM-based system \rev{designed specifically to help PVI with} visual interpretation and question answering\rev{. Built on} OpenAI's GPT-4 models~\cite{achiam2023gpt}, \sbma{} offers \rev{capabilities that surpass those of} similar applications.

Prior work~\cite{adnin2024look} has investigated \rev{how PVI use and access generative AI tools,} focusing on information access, acquisition, and content creation through platforms like \bma. Our work \rev{broadens this investigation beyond} content-oriented interactions to \rev{examine} real-life scenarios \rev{involving} visual descriptions, task performance, social interactions, and navigation. By \rev{analyzing} the capabilities and limitations of LMM-based VQA systems, particularly \bma{}, we identify gaps between \rev{current} technological capabilities and \rev{PVI's} practical needs and expectations. 
This study offers timely insights into rapidly evolving LMMs \rev{while seeking broader understanding that will remain valuable as foundational knowledge even as specific technologies advance.} 

\rev{
More specifically, we explore} the following \textit{research questions}:
\begin{itemize}
    \item \textit{What are the capabilities and limitations of LMM-based assistance in the daily lives of people with visual impairments?}
    \item \textit{How do people with visual impairments mitigate these limitations?}
\end{itemize}

To answer these questions, we conducted an exploratory study \rev{using two complementary data sources. First, we interviewed} 14 visually impaired users \rev{about their experiences with \sbma. This was our primary data source. Second, we analyzed image descriptions generated by this tool that were shared by both our interview participants and users on} social media platforms (X, Facebook, Reddit, and blogposts). This \rev{secondary data source allowed us to understand PVI's lived experiences with this AI tool while capturing concrete examples of their interactions in various real-life scenarios.}

\rev{Our study revealed} that \sbma's context-aware capabilities \rev{help participants better understand their} spatial \rev{surroundings, support} social \rev{interactions, interpret} stylistic \rev{elements, and convey} human identities. \rev{Yet} several limitations \rev{undermine} these benefits\rev{: the AI sometimes hallucinates} non-existent details, \rev{makes} subjective interpretations \rev{about} human-animal interactions and fashion \rev{choices, and misidentifies} people's age or gender. \rev{When encountering these limitations, participants draw on their spatial memory and auditory cues, apply personal judgment, or turn to human assistants.} Moreover, \rev{\sbma{} struggles to grasp users' intentions, provide} actionable support\rev{, or offer} real-time feedback. Participants compensate for this by actively guiding it through prompting, seeking human \rev{assistance}, or depending on their orientation and mobility skills.

Informed by these findings, we discuss strategies to improve handoff between PVI, the AI tool, and remote sighted assistants. We propose streamlined interactions to reduce redundancy and envision new paradigms to achieve more accurate identity recognition, improve subjective interpretations, and mitigate AI hallucinations. 
We also discuss the benefits of multi-agent systems, including both human-human and human-AI collaborations, and explore future possibilities of AI-AI cooperation to aid PVI in tasks requiring specialized knowledge.

\rev{
Our research examines how state-of-the-art AI, particularly} LMM-based systems like \sbma, \rev{creates new opportunities for PVI through advanced human-like language and vision capabilities. Rather than contrasting this tool with other VQA systems, we characterized this as a new genre of AI-driven prosthetic and identified how its early-stage usage could better align with established assistive approaches such as remote sighted assistance~\cite{lee2020emerging,kamikubo2020support,granquist2021evaluation}. 
Through detailed analysis of an AI tool's capabilities and limitations, based on rich narratives from PVI about their first-hand experiences, we provide insights into how LMM-based systems are reshaping accessibility tools. These insights help us understand how to enhance both the context awareness and intent-oriented capabilities of LMMs, laying the groundwork for future advances in intelligent, interactive, and personalized assistive technology that better serve PVI's needs.
}

\section{Background and Related Work}


In this section, we review the literature on AI-powered and human-assisted visual interpretation and question answering systems, as well as information needed in visual interpretations for PVI.

\subsection{AI-Powered Visual Interpretation and Question Answering Systems}

The advancements in deep learning models for computer vision and natural language processing have significantly enhanced the capabilities of AI-powered visual assistive systems\rev{~\cite{ahmetovic2020recog, mukhiddinov2022automatic}}. These systems leverage photos taken by PVI to identify objects or text within the scene, as well as respond to queries about the contents of the images\rev{~\cite{hong2022blind, morrison2023understanding}. Such applications are now widely adopted, exemplified by Microsoft's Seeing AI~\cite{SeeingAI2020}, providing support to PVI in various scenarios~\cite{gonzalez2024investigating}.}


Recent advancements in LMMs, such as GPT-4~\cite{achiam2023gpt}, have demonstrated exceptional performance in multimodal tasks~\cite{yu2023mm}, \rev{prompting exploration into their potential for assisting PVI. State-of-the-art LMMs have been leveraged to create assistive systems capable of evaluating image quality, suggesting retakes, answering queries about captured images~\cite{zhao2024vialm, yang2024viassist}, and even integrate multiple functions to assist PVI in tasks such as navigation~\cite{zhang2025enhancing} and text input~\cite{10.1145/3613904.3642939}. In the commercial domain, Be My Eyes~\cite{BeMyEyes2020} introduced the \bma{} feature powered by GPT-4~\cite{achiam2023gpt}.}



Prior work has examined the use of AI-powered visual assistive systems by PVI in their daily lives\rev{~\cite{granquist2021evaluation,kupferstein2020understanding,gonzalez2024investigating}}. However, these studies did not incorporate the latest LMM technologies, thus may not fully represent the user experience of cutting-edge AI-powered systems. Bendel~\cite{bendel2024can} documented his experience with GPT\mbox{-}4\mbox{-}based \bma, yet his account remains subjective. Therefore, in-depth investigation into PVI's daily utilization of state-of-the-art AI-powered systems is imperative. 
This paper addresses this gap by exploring how 14 visually impaired users incorporate \bma, a state-of-the-art LMM-based VQA system, into their daily routines.




\subsection{Human-Assisted Visual Interpretation and Question Answering Systems}

Human-assisted VQA systems offer prosthetic support for PVI by facilitating connections to sighted people through remote assistance. These systems utilize image-based and video-based modalities to \rev{meet varied} needs and situations.

Image-based human-assisted VQA systems allow PVI to submit photos along with their queries and receive responses after some time. An example of this is Vizwiz, where PVI can upload images accompanied by audio-recorded questions and receive text-based answers through crowdsourced assistance~\cite{bigham2010vizwiz_nearly}. This method has been successfully applied in tasks such as reading text, identifying colors, locating objects, obtaining fashion advice, \rev{and supporting social interactions}~\cite{bigham2010vizwiz_nearly, bigham2010vizwiz, burton2012crowdsourcing,gurari2018vizwiz}. 
However, the single-photo, single-query limitation of image-based VQA systems~\cite{bigham2010vizwiz_nearly} makes it less suitable for addressing complex or contextually deep inquiries~\cite{lasecki2013answering}.

Conversely, video-based human-assisted VQA systems facilitate real-time, interactive support, allowing PVI to receive immediate assistance tailored to their specific environmental context. This approach enables the visual interpretation of real-time scenes and supports a dynamic, back-and-forth VQA process, which is essential for addressing more specific and complex contextual inquiries effectively.
The evolution of this technology has progressed from wearable digital cameras ~\cite{hunaiti2006remote,garaj2003system,baranski2015field} and webcams~\cite{bujacz2008remote,scheggi2014remote,chaudary2017tele} to the utilization of mobile video applications~\cite{holmes2015iphone, BeMyEyes2020, Aira2020,xie2022dis,xie2024bubblecam} and real-time screen sharing technologies~\cite{lasecki2011real,lasecki2013answering,xie2023two}. 
Services like Be My Eyes~\cite{BeMyEyes2020}, which connects PVI with untrained volunteers, and Aira~\cite{Aira2020}, which connects PVI with trained professional assistants, exemplify the application of video-based VQA in scenarios that require immediate feedback. These services prove effective in navigation~\cite{kamikubo2020support,xie2022dis,c4vtochi,iui,yu2024human}, 
shopping~\cite{c4vtochi,xie2023two,iui}, 
and social interaction~\cite{lee2020emerging,Caroll2020Human,lee2018conversations}. 
\rev{
In this study, we revealed specific situations where participants used human-assisted VQA systems to address limitations in \bma's assistance.
}

\subsection{Information Needed in Visual Interpretations for Visually Impaired Users}

\rev{Both} AI-powered \rev{and} human-assisted visual interpretation face the challenge \rev{of identifying what specific information is necessary to meet PVI needs}. Existing research primarily focuses on \rev{what information PVI seek in online image descriptions}. The Web Content Accessibility Guidelines~\cite{caldwell2008web} \rev{advise on creating alternative text (alt text), but their one-size-fits-all approach limits context-specific applicability}. Similarly, AI-based \rev{tools like Seeing AI~\cite{SeeingAI2020} adhere to this uniform design model.}

\rev{Earlier studies~\cite{gurari2018vizwiz, gurari2020captioning} relied on} sighted volunteers \rev{to decide image descriptions for PVI, while recent research~\cite{stangl2020person,stangl2021going,bennett2021s,chen2024role} emphasizes} user-centered approaches. For example, Stangl et al.\cite{stangl2020person} \rev{found PVI preferences for descriptions vary by platform} (e.g., news, social media, eCommerce)\rev{, with both common needs (e.g., identifying gender, naming objects) and platform-specific details (e.g.,} hair color on dating websites). \rev{Follow-up work~\cite{stangl2021going} showed preferences also depend on users' goals, with subjective details rarely included.} Bennett et al.~\cite{bennett2021s} examined screen reader users' \rev{views on describing} race, gender, and disability \rev{in appearance descriptions.}
\rev{For video content,} Jiang et al.~\cite{jiang2024s} \rev{examined PVI preferences across genres (e.g., how-to, music videos), recommending that entertainment descriptions focus on subjects, while how-to videos prioritize actions and tools.} Natalie et al.~\cite{natalie2024audio} \rev{studied PVI needs for customization in video descriptions, such as length, speed, and voice. For synchronous systems like real-time video interpretation,} Lee et al.~\cite{lee2020emerging} identified \rev{PVI needs} in remote sighted assistance\rev{, including} both objective details (e.g., text, spatial information\rev{) and subjective input} (e.g., opinions on clothing).
In this study, we will examine how the LMM-based system processes visual information and provide design insights based on the information needed by PVI.

\rev{




}

\section{Method: Data Collection and Analysis}

Our exploratory study used qualitative methods to examine how PVI use \bma, an emerging LMM technology, in their daily lives.
This section details our data collection and analysis processes.

\subsection{Data Collection}

\rev{
We gathered data from two sources: interviews as the primary data source and \sbma-generated image descriptions as the secondary data source. 
Interview data captured accounts of PVI's lived experiences with \bma, enabling participants to articulate personal interactions, challenges, and perceptions of the technology. 
The image descriptions enriched and supplemented interview data by providing concrete examples of how PVI used this tool across various contexts, from daily chores to professional tasks and travel.

We collected the two data sources sequentially. 
First, we interviewed $14$ visually impaired users, focusing on \sbma's emerging practices in their daily lives. 
At the end of each interview, we invited participants to share their image descriptions. 
However, most participants didn't have examples to share or didn't want to share due to the demanding process or personal content sensitivity (e.g., images capturing their faces). Respecting and prioritizing participant privacy, we ensured voluntary sharing of image descriptions. Consequently, 4 participants shared their descriptions voluntarily.
Meanwhile, we accessed and analyzed publicly available experiences shared by PVI on reliable online platforms.

The combination of text-based interview data and text/visual image descriptions provided richer insights into PVI's experiences with emerging LMM technologies than either source alone would have offered. We detail our data collection process in the following section. 
}

\subsubsection{Primary Data Source: Interview Data}
We recruited 14 visually impaired participants (10 blind, 4 low-vision; 4 males, 10 females) through our prior contacts and snowball sampling.  
All participants actively used \bma{} and were primarily aged 35-40. Their occupations included three students and nine employed individuals, with two participants unemployed. 
Table~\ref{user_demographic_info} presents their demographics. 
Each participant received a \$30 gift card per session for their time and effort.

\begin{table*}[]
\small
\caption{Participants' demographics.}
\label{user_demographic_info}
\begin{tabular}{p{0.3cm}p{0.8cm}p{0.8cm}p{4.8cm}p{2.8cm}p{3cm}p{2.1cm}}
\toprule
\textbf{ID} & \textbf{Gender} & \textbf{Age Group} & \textbf{Condition of Vision Impairment} & \textbf{Age of Onset} & \textbf{Occupation Type} & \textbf{Be My AI Usage Frequency} \\ \toprule
P1 & F & 45-50 & Totally blind, retinopathy of prematurity & Since birth & IT consultant &  3 or 4 times a day \\ \hline
P2 & F & 35-40 & Low vision, cone-rod dystrophy & Since birth & Program director in a nonprofit &  a few times a week \\ \hline
P3 & F & 30-35 & Totally blind, Leber's congenital amaurosis & Since birth & Elementary school teacher & 5 times a week \\ \hline
P4 & M & 25-30 & Totally blind, Pale optic nerves & More than 12 yrs ago & Criminal law employee & 2 to 3 times a week \\ \hline
P5 & F & 40-45 & Totally blind, retinopathy of prematurity & Since birth & Manager of digital accessibility & 2 times a day \\ \hline
P6 & F & 25-30 & Totally blind, microcephaly and detached retina & Since birth & Student & once a week \\ \hline
P7 & F & 35-40 & Low vision, retinopathy of prematurity & Since birth & Part-time employee &  2 times a week \\ \hline
P8 & M & 40-45 & Totally blind, detached retina & Since birth & Insurance & a few times a day \\ \hline
P9 & F & 30-35 & Totally blind, retinopathy of prematurity & Since birth & In-between jobs & a few times a week \\ \hline
P10 & F & 30-35 & Low vision, retinitis pigmentosa & Since birth & Student &  3 or 4 times a day \\ \hline
P11 & M & 35-40 & Totally blind, retinopathy of prematurity & Since birth & Stay-at-home parent &  3 or 4 times a day \\ \hline
P12 & M & 20-25 & Low vision, Leber's hereditary optic neuropathy & Since 14 yrs old & Student &  2 times a week \\ \hline
P13 & F & 40-45 & Totally blind, retinitis pigmentosa & Low vision since infancy, totally blind since 2021 & Human service employee &  5 times a week \\ \hline
P14 & F & 35-40 & Totally blind, retinopathy of prematurity & Since a few months old & Assistive technology specialist &  5 times a week 
\\ \bottomrule
\end{tabular}
\Description[Blind participants' demographic information about gender, age, condition of vision impairment, occupation type, and Be My AI usage frequency]{Ten blind participants are female, and four blind participants are male. Their age group ranges from 25 to 50, and their occupations include student, IT consultant, stay-at-home parent, teacher, employees in nonprofit and insurance, and assistive technology specialist. Their frequency of Be My AI usage varies from 2 times a week to 4 times a day.}
\end{table*}

\paragraph{Procedure} We conducted individual semi-structured interviews via Zoom lasting 50-76 minutes, with one or two researchers present per session. All interviews were recorded with participant consent. 

The interviews followed four main phases. 
\textit{First}, we invited participants to share their personal use cases for \bma. 
To facilitate recall, we referenced Be My Eyes' common use case list~\cite{bma_usecase}, prompting participants to discuss similar experiences. 
Follow-up questions were posed to further investigate their experiences with each identified use case. 
\textit{Second}, participants evaluated the quality of the tool's visual interpretations, focusing on accuracy, detail level, error, and appropriateness of identity descriptions. 
\textit{Third}, we explored how participants used \sbma{} among their various assistive tools, including human-assisted and AI-powered VQA systems, to understand its distinct advantages and limitations. 
\textit{Finally}, we collected demographic information and inquired about participants' willingness to share copies of its visual interpretations for research analysis.

\subsubsection{Secondary Data Source: Image Descriptions Generated by \bma}
Image descriptions collected from interview participants and social media platforms served as a secondary data source to complement our interview findings. These descriptions included either text-only copies, or original images sent to \sbma{} with their corresponding descriptions. Some descriptions also included follow-up questions and responses between users and the system.

\paragraph{From Participants}

Participants were given the flexibility to select their preferred documentation method, either copying text or taking screenshots. 
In total, we gathered 22 image descriptions from 4 participants, as most participants either lacked examples or were reluctant to share due to privacy concerns or the effort required in documentation.
Of these descriptions, 4 were original images alongside their descriptions, while the majority consisted of text copies -- a format participants found more manageable. 
Additionally, 6 of the 22 descriptions contained follow-up questions where participants sought clarification or additional details from the system.

\paragraph{From Social Media Platforms}

We collected image descriptions from 4 social media platforms: X\footnote{\url{http://x.com/}}, Facebook\footnote{\url{https://www.facebook.com/}}, Reddit\footnote{\url{https://www.reddit.com/}}, and blogposts. The platforms were chosen to cover both image-sharing sites (X, Facebook) and discussion forums (Reddit, blogposts).

We used the search terms ``BeMyAI'' and ``\#bemyai'' to find relevant posts on X, Facebook, and blogposts. For Reddit, we searched for ``AI'' and ``BeMyAI'' within the r/Blind community.

We gathered posts that met the following criteria: (i) written in English, (ii) published between September 25, 2023, the official release date of \bma, and March 31, 2024, and (iii) contained image descriptions generated by the tool. 
To avoid redundancy, if identical content appeared across multiple platforms, only the earliest published post was recorded. 
If one post included multiple descriptions, each description was documented as a separate entry. 
If there were multiple posts published around the same time frame about the same topic (e.g., an X or Reddit thread), they were treated as a single entry.

We verified the authenticity of posts by confirming that they were from either the official Be My Eyes account or from users who identified themselves as visually impaired in their profiles. These established platforms serve as reliable sources where PVI regularly share their experiences.

In total, we collected 28 \sbma-generated image descriptions from 4 social media platforms. Of these, 23 included original images sent to \sbma{} \rev{and 5 were text-only interactions.} Fifteen descriptions contained follow-up questions that participants sent to the system for additional details or clarification. 
By platform, we collected 
(i) X: 17 image descriptions (16 with original images, 6 with follow-up questions), 
(ii) Facebook: 3 image descriptions (all with original images), 
(iii) Reddit: 3 image descriptions (1 with follow-up questions), and 
(iv) blogposts: 5 image descriptions (4 with original images, 8 with follow-up questions). 
The complete dataset is available at \href{https://bit.ly/4hqo5ve}{\textit{\textcolor{black}{https://bit.ly/4hqo5ve}}}.

\rev{This dataset, although limited in size, captures crucial insights into early user experiences with \bma. By collecting descriptions shortly after its release, we documented initial user interactions and system performance during its early phase. 
}

\subsection{Data Analysis}

We used a bottom-up approach \rev{to analyze the interview data}. The first author conducted inductive thematic analysis~\cite{braun2006using} by developing initial codes through open coding, then iteratively collating and grouping these codes into themes and categories. All authors reviewed and finalized these themes during weekly meetings (see Table~\ref{codebook} in Appendix).

\rev{
We used a top-down approach to analyze the image descriptions. 
The first author examined each image description alongside its contextual elements: the user's follow-up questions, \sbma's responses, and the circumstances of image capture. Using the codebook derived from our interview analysis, we then analyzed these descriptions deductively.

The following example illustrates our data analysis process from inductive to deductive approach.
During interviews, participants mentioned instances where \sbma{} initially provided general scene descriptions rather than their desired specific content, requiring participants to guide it through follow-up questions. We labeled this interaction as ``Goal Understanding Dialogue.'' 
When analyzing image descriptions, we identified similar patterns (Figure~\ref{eggshells}) and classified them under this pre-established ``Goal Understanding Dialogue'' category.

Through this process, we first identified the system's capabilities and limitations through interview data, then complement these interview findings with specific examples from the image descriptions and subsequent conversations between users and the system.

}

\section{Findings}
\label{findings}

In this section, we first explore \sbma's context-aware capabilities across diverse settings, assessing its effectiveness and constraints in interpreting physical environments, social and stylistic cues, and people's identity. 
Next, we examine its intent-oriented capabilities by evaluating both the strengths and limitations of the system's ability to understand and act on users' intentions. 
For each aspect, our findings unfold in two parts. In \textit{Capabilities of \bma}, we examine how PVI use the system and reveal its capabilities. In \textit{Example}, we illustrate specific scenarios that present both the challenges users faced and their adaptive strategies.

\subsection{Context Awareness}

This section examines how \bma{} enhances user experiences across diverse environments and interactions. We analyze its roles in enhancing spatial awareness in physical settings, interpreting social and stylistic context, and conveying human identities. 
Our analysis considers both the system's strengths in providing context-rich descriptions and its limitations due to technical constraints and subjective interpretations.

\subsubsection{Physical Environments}
\label{physical_environments}

Participants used \sbma{} to enhance their perception of indoor and outdoor environments through detailed scene descriptions. 
They explored various settings, from theaters (P2) and room layouts (P3, P7, P9, P13) to holiday decorations (P5) and street views (P11, P12). While the system provides structured visual information that expands spatial awareness, its effectiveness occasionally diminishes due to AI hallucinations or requires participants' guidance for accurate scanning.

\paragraph{Capabilities of \bma} Seven participants (P2, P3, P7, P9, P11, P13, P14) reported that the AI tool's comprehensive scene descriptions enhanced their spatial awareness. 
These descriptions often uncover spatial details previously unnoticed by users, as noted by P11, \textit{``It gives me unexpected information about things that I didn't even know were there.''} 
%

However, this enhanced awareness is compromised by AI hallucinations, instances where AI incorrectly identifies nonexistent objects. 
Such errors can disrupt the user's context understanding, leading to confusion or misinterpretation of the environment. To verify \sbma's descriptions, participants either seek human assistance or rely on their pre-existing mental models of the environment.

Participants successfully guide \sbma{} using auditory cues, particularly when locating dropped objects. They combine their hearing and spatial memory to adjust camera angles, which enables the tool to scan more accurately. 
This integration of participants' auditory inputs guides and refines the system performance, leading to more accurate and useful descriptions.

\paragraph{Example 1: AI Hallucinations of Adding Non-existent Details}

Be My AI provides detailed descriptions of objects, characterizing their colors, sizes (e.g., ``small,'' ``medium,'' ``large,'' and ``tall''), shapes (e.g., ``round,'' ``square,'' ``fluffy,'' ``wispy,'' ``dense,'' and ``open lattice structure''), and spatial orientations (e.g., ``on the left,'' ``to the right,'' ``in the foreground,'' ``in the middle,'' and ``in the background''). 
However, four participants (P3, P5, P11, P14) encountered AI hallucinations where the tool fabricated details in their environments.

Participants developed two strategies to verify the accuracy: consulting human assistants and drawing on their spatial memory.
P3's experience illustrates the value of human verification. When \bma{} reported a \textit{``phantom''} object behind her, she first investigated personally, finding nothing. A human assistant then confirmed the absence of any object. 

\begin{quote}
    \textit{``Apparently, Be My AI said there was an object behind me, and there really wasn't. So, then, when I went and asked somebody, `You know what is behind me?' and they said there wasn't anything behind you in the picture.''} (P3)
\end{quote}

P5's experience demonstrates how spatial memory helps identify inaccuracies. When the system incorrectly described objects next to her dogs in her home, her familiarity with the space enabled her to recognize these errors independently. This ability to leverage environmental knowledge allows participants to detect and disregard AI hallucinations effectively.

\paragraph{Example 2: Need for User Support in Locating Dropped Objects}

Participants (P4, P12, P13) used \sbma's sequential \textit{``top to bottom, left to right''} descriptions to locate dropped items like earbuds and hair ties. 
These descriptions provided precise locations, such as when P13 learned of \textit{``a picture of a carpet with a hair tie in the upper right hand corner,''} and when P12 discovered \textit{``the earphones are directly in front of you, between your feet.''}
Such sequential descriptions allow participants to pinpoint lost objects with greater accuracy.

To optimize the tool's scanning effectiveness, participants first used their auditory perception and spatial memory to approximate an object's location before positioning the camera. 
This adjustment allowed the tool to focus on the intended search area, rather than random searching. 
For instance, P4 utilized his \textit{``listening skills''} to detect the location of a fallen object before directing the camera to that specific spot. Similarly, P12 enhanced the camera's view of dropped earbuds by stepping back from his seat to capture a better angle of the floor. 
These examples illustrate how participants synergize their understanding of the environment with the system's technological capabilities to manage tasks that require spatial awareness.

\subsubsection{Social and Stylistic Contexts}
\label{social_stylistic}

Participants leveraged \sbma{} for both social interactions and style-related decisions. 
In social settings, nine participants (P2, P4, P5, P8-11, P13, P14) used it to take leisure pictures of animals like cats, dogs, birds, and horses, to better understand and monitor animal behavior and status. 
In stylistic applications, it aids in identifying clothing colors (P2, P7, P9, P13), recognizing patterns (P2, P9), coordinate outfits by matching clothes with shoes and jewelry (P2, P4, P5, P6, P12), and assessing makeup (P5, P10).
This section examines the system's subjective interpretations within these contexts, highlighting its utility in enriching users’ experiences and the challenges in providing accurate, context-aware descriptions.

\paragraph{Capabilities of \bma}

\sbma{} usually concludes its descriptions by injecting subjective interpretations, which enriches users' understanding of depicted scenarios. 
These interpretations encompass people's or animal's emotional states (e.g., ``a cheerful expression on his face,'' ``smiling slightly,'' ``friendly and approachable,'' ``curious expression''), 
body language (e.g., ``his arms are open as if he is engaging in a conversation or greeting the other man''),
and the ambiance (e.g., ``peaceful and natural,'' ``serene and peaceful atmosphere,'' ``cozy and cheerful holiday vibe,'' ``warm and inviting atmosphere,'' ``the atmosphere seems to be lively and festive'').

\begin{figure}[t!]
\centering
\includegraphics[width=0.4\textwidth]{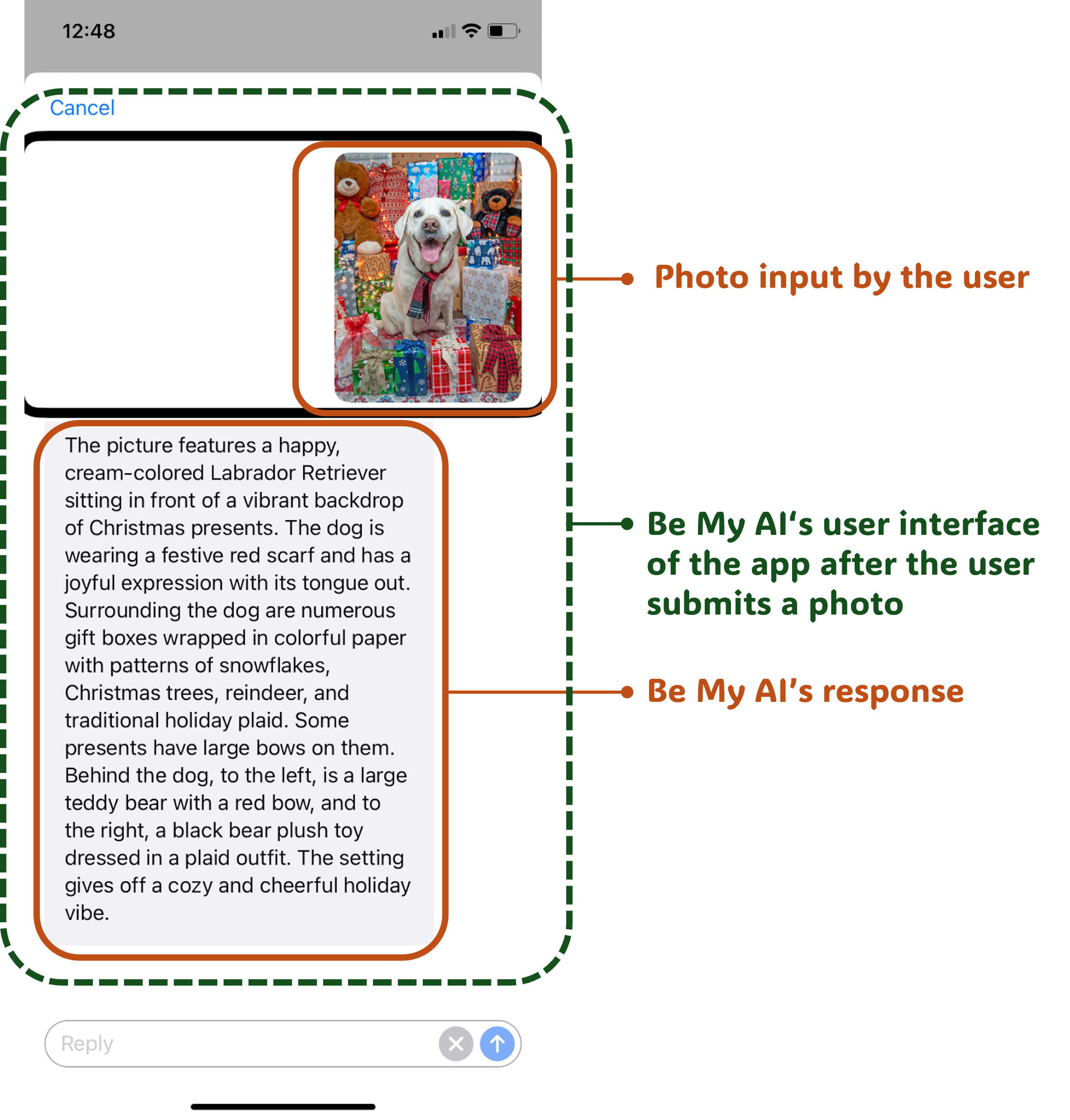}
\caption{\bma's description of a dog, including subjective interpretations of the dog's emotions. This screenshot was provided by P9.}
\Description{Interface of Be My AI. Top of the screen is a photo of a dog. Below the photo is the description of the dog, including subjective interpretations of the dog's emotions. This screenshot was provided by P9. The full description is as follows. The picture features a happy, cream-colored Labrador Retriever sitting in front of a vibrant backdrop of Christmas presents. The dog is wearing a festive red scarf and has a joyful expression with its tongue out. Surrounding the dog are numerous gift boxes wrapped in colorful paper with patterns of snowflakes, Christmas trees, reindeer, and traditional holiday plaid. Some presents have large bows on them. Behind the dog, to the left, is a large teddy bear with a red bow, and to the right, a black bear plush toy dressed in a plaid outfit. The setting gives off a cozy and cheerful holiday vibe.}
\label{human_animal}
\end{figure}

These subjective interpretations help participants better engage in social interactions and understand animal behavior. 
Participants (P5, P9-11) specifically used the system to grasp nuances in animals' facial expressions, activities, and body language (Figure~\ref{human_animal}). P10 highlighted its utility during dog walks: \textit{``Sometimes it's hard for me to know if the dog is peeing or what the dog is doing.''}

Furthermore, the subjectivity extends to fashion suggestions, with \bma{} recommending stylistically coordinated outfits based on colors and patterns, as well as assessing makeup. 
For instance, P5 employed it to check the color, placement, and overall balance of her makeup, P2 and P9 consulted it to coordinate tops and bottoms, and P12 used it to select a tie that complemented his shirt.

However, participants noted that BMA's visual interpretations sometimes included excessive subjective elements. 
This subjective input, unverified by human, could potentially undermine the system's ability to provide accurate context awareness.  
As a result, participants preferred their own subjective interpretations or feedback from sighted assistants in contexts involving human-animal interactions and fashion choices.

\paragraph{Example 1: Subjective Interpretations in Human-Animal Interactions}

Participants (P2, P5) identified limitations in how \sbma{} infers animal emotions in its image descriptions, raising concerns about the accuracy of these subjective interpretations. 
For instance, P5 highlighted instances where the tool describes a dog \textit{``appears to be relaxed''} or \textit{``appears to be happy.''} 
Likewise, P2 noted its tendency to include subjective commentary, as in \textit{``That's a white cat curled up on a fuzzy blanket. She looks peaceful and happy and rested.''}

These interpretations go beyond observable visual elements to make emotional inferences that may not reflect reality. 
As a result, participants expressed a preference for objective, fact-based descriptions. 
P2 articulated a desire for less editorializing, saying, \textit{``Maybe I don't want it to editorialize, you know, maybe I just literally only want the facts of it.''}

Participants (P2, P5) stressed the importance of maintaining human agency in interpreting animal behaviors, preferring their own judgment rather than the system's subjective interpretations. 
P5 particularly valued the ability to modify \sbma's descriptions of her dog's expressions, maintaining control over her interpretation of pet behaviors.

\begin{quote}
    \textit{``Some blind people think, `How does it know that the dogs are happy? Why does it assume?' Some people don't like that it's making assumptions about the picture. I like having access to that information, but I like to be able to change it if I want.''} (P5)
\end{quote}

In summary, \sbma's subjective inferences can enrich descriptions, yet they risk introducing inaccuracies that undermine the system's reliability. Users therefore value the ability to override these interpretations, preserving their agency in understanding the context.

\paragraph{Example 2: Subjective Interpretations in Fashion Help}

\sbma's subjective interpretations extend beyond human-animal interactions to fashion help. 
Participants (P6, P7, P10) utilized it to describe colors and patterns but expressed concerns about its subjective fashion suggestions. 
They preferred to make their own style choices or seek human assistance for outfit matching, highlighting the human subjectivity in fashion decisions.

P6 questioned the AI's capacity to authentically replicate human judgment, saying, \textit{``It's interesting how AI is being taught to simulate kind of the human factor of things.''}
She cited experiences where AI-generated responses appeared \textit{``strange''} and \textit{``complete nonsense,''} contrasting these with the nuanced understanding that humans provide.

\begin{quote}
    \textit{``No, no, no, no, I would never use it to do anything that required human subjectivity... I just don't trust AI with a task that is supposed to be subjective like that, particularly visual like that. Have you ever seen AI weirdness?... I think that just goes to show why I'm not gonna trust AI with my fashion yet.''} (P6)
\end{quote}

P10 explicitly preferred human feedback, stating, \textit{``I'm still more confident asking a sighted person to provide me with the feedback.''} 
This preference underscores how participants maintain their agency by relying on human judgment for fashion decisions rather than deferring to \sbma's suggestions.

\subsubsection{Identity Accuracy and Sensitivity}
\label{identity}

Nine participants (P2, P3, P5-7, P9-11, P13) evaluated the system's ability to describe people's identities in images of their families, friends, and social media posts. 
Our analysis examines how the tool handles identity attributes like age and gender, focusing on both its capabilities and limitations in providing sensitive and accurate descriptions.

\paragraph{Capabilities of \bma}

\sbma{} describes people's identity in terms of gender (e.g., ``woman,'' ``man''), age (e.g., ``old,'' ``late twenties or early thirties''), appearance (e.g., ``long, dark brown hair,'' ``wavy brown hair,'' ``His hair falls past his ears, with a slightly messy but stylish look'') and ethnicity (e.g., ``East Asian''). 
Participants appreciated such detailed descriptions for providing deeper insight into people's visual characteristics. P2 explained: \textit{``I enjoyed hearing, I knew my friend was East Asian when I saw her picture. It was kind of cool to see that because... as a blind person, you don't know that all the time. So, I'd like access to it.''}

However, eight participants (P2, P3, P5-7, P9, P10, P13) encountered challenges with the system's accuracy and sensitivity in describing identity attributes, particularly gender and age. 
These errors reveal the tool's limited ability to interpret contextual cues that extend beyond visual appearance, such as cultural, situational, and personal contexts.

\paragraph{Example: Inaccurate Identification of People's Gender and Age}

Three participants (P3, P10, P13) reported inaccuracies in \sbma's gender identification. 
P13 described how it misidentified gender by mistaking a hidden ponytail for short hair: \textit{``just looked like short-cropped hair.''} 
Similarly, P10 revealed that the system struggled with gender identification for individuals whose physical attributes do not conform to typical gender norms, noting, \textit{``the person had a short hair, and was a female''}. 
In response, it defaulted to neutral language, \textit{``\bma{} didn't tell me it was a woman or a man, just said a person.''}
These examples illustrate the system's reliance on stereotypical indicators and reveal its challenges in interpreting non-visible details.

Additionally, two participants (P5, P9) identified problems with age description. P9 reported that \bma{} overestimated her daughter's age by one year. 
Likewise, P5 observed sensitivity concerns when the system labeled someone as \textit{``an old woman''} based solely on grey hair. 
To address these age-related inaccuracies, P5 suggested that the tool should adopt more objective, factual descriptions rather than subjective or potentially stigmatizing labels, such as \textit{``a woman with grey hair, instead of an older woman.''}
These examples indicate the system's difficulties with precise age estimation and the potential inaccuracy when AI makes assumptions based on appearance alone.

In summary, these examples illustrate the tool's limited context awareness and its challenge in accurately interpreting people's identities, such as gender and age, due to the reliance on stereotypical visual cues. These misjudgements are rooted in the system's inability to integrate and interpret broader, non-visible contextual elements like cultural norms and personal styling choices.

\subsection{Intent-Oriented Capabilities}

In this section, we explore \sbma's limitations \rev{in} comprehending users' goals, providing actionable support to fulfill these objectives, and offering real-time feedback. 
Through analysis of specific examples, we elucidate the extent of the system's intent-oriented capabilities and highlight areas where it falls short in adapting to user needs.

\begin{figure}[t!]
\centering
\includegraphics[width=0.45\textwidth]{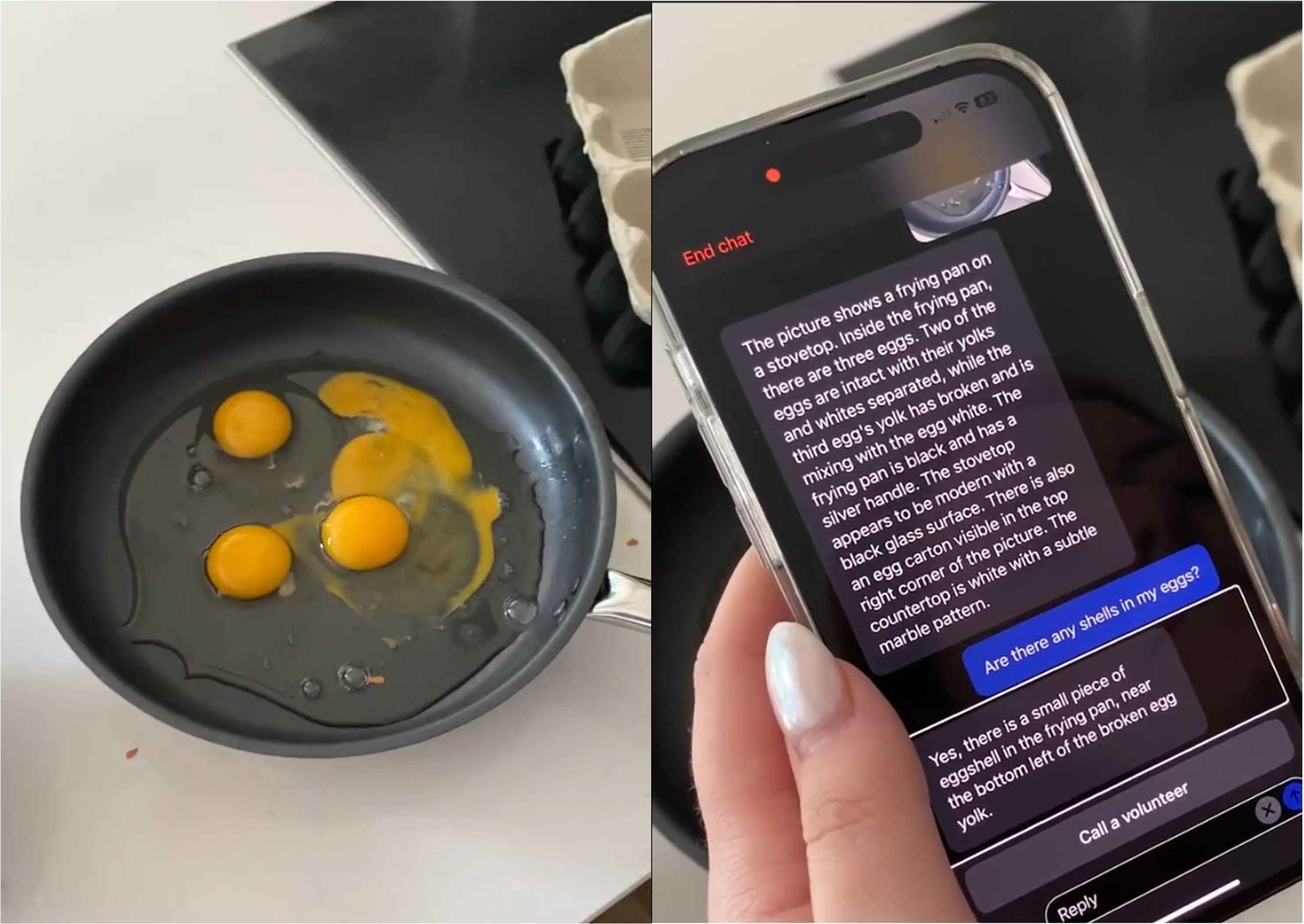}
\caption{On the left is the original image sent to \bma. On the right is \bma's description of eggs in a frying pan, followed by a question checking for the presence of eggshells. This example was originally drawn from X.}
\Description{On the left is the image of a frying pan on the stove-top. Inside the pan there are 4 eggs, one of whose yolk is broken. On the right is a photo of an user using Be My AI on a smart phone.}
\label{eggshells}
\end{figure}

\subsubsection{Agentic Interaction}
\label{agentic_interaction}

While \bma{} can process visual information, it often struggles to infer users' specific intentions from images. Users compensate for these limitations by guiding the system through targeted prompting.

\paragraph{Capabilities of \bma}

BMA's ``ask more'' function enables users to explore image details that were not covered in initial descriptions. Through this feature, users can probe for specific details that match their personal needs or interests.
Eleven participants (P1-3, P5, P6, P8-12, P14) used the ``ask more'' function to 
match outfits (P2, P5, P12),
check makeup (P5, P10), 
suggest cooking recipes (P12), 
assist with household appliances (P5, P12), 
examine text or objects (P6, P8, P11, P12, P14),
gain more details about people's facial expressions, attire, or actions (P2, P3, P5, P8, P9, P10, P11), 
check animal status (P3, P5, P11, P14), 
and edit the descriptions for social media posts (P2, P3).

Seven participants (P1-3, P5, P8-10) valued the flexibility of posing follow-up questions, enhancing their independence by reducing reliance on human assistance. As P3 noted, \textit{``I can ask follow-up questions, so I have a good way to sort of figure out what's in the image independently, which is something I was not able to do before this app came out.''}

However, participants identified a common issue that the system is unable to discern users' goals in initial response, often resulting in generalized descriptions without knowing which aspects to emphasize. P12 elaborated on this challenge, stating, \textit{``\bma{} loves to make general descriptions, and it doesn't know what to focus on.''} 
To overcome this limitation, users guide the system with targeted prompts that clarify their specific intentions.

\paragraph{Example 1: Check for the Presence of Eggshells}

In Figure~\ref{eggshells}, BMA's initial response provides a detailed descriptions of what is included in a frying pan. 
The user, likely influenced by prior experiences of inadvertently leaving unwanted items in the pan while cooking, specifically intended to check for such elements. 
However, the system did not infer that the user was looking for unwanted elements like eggshells, rather than seeking a general description of the scene.
To clarify her objective, the user posed a follow-up question, thus directing \sbma{} to recognize and prioritize her specific concerns and intentions.

In this example, although the AI tool can describe the visual elements accurately and comprehensively (from the condition of the eggs to the surroundings), it failed to identify unusual visual elements that are challenging for PVI to detect but are important to their understanding of the scene. 
User prompting helps refocus the system's attention on these unusual elements, thereby enhancing its ability to interpret and respond to user-specific intents more effectively.

\paragraph{Example 2: Adjust Rotary Control Appliances}
Six participants highlighted the tool's inadequacy in comprehending their goal of adjusting rotary control appliances like washers (P1) and thermostats (P2, P4, P5, P13, P14). This task requires \bma{} to interpret the current settings on these appliances as users modify dials for time, mode, or temperature.
However, the system often offered broad descriptions of visual elements without honing in on the user's specific objectives.

P13's experience exemplifies this limitation. While the system could recognize a thermostat on the wall, it failed to provide critical information such as the current temperature setting or instructions for adjusting the temperature.

To overcome the limitation, participants guided the system to better understand their goal through prompting. 
For instance, P5 asked precise questions like \textit{``What is the arrow pointed at right now on the current setting?''} Such specific queries helped redirect \sbma's focus from general scene descriptions to the exact details needed for appliance adjustment.

\subsubsection{Consistency and Follow-Through}
\label{consistency}

In Section~\ref{agentic_interaction}, we described how users needed to explicitly guide \bma{} when it failed to understand their intentions. 
This section examines the system's difficulties maintaining consistency and following through on tasks even after acknowledging users' goals.

\paragraph{Capabilities of \bma}

\sbma{} often fails to proactively suggest next steps or support users until they achieve their objectives. 
Effective consistency requires maintaining focus on the user's objective throughout the interaction. \rev{This includes} offering progressively specific and relevant assistance, and ensuring all responses contribute to the user's intended outcome.

Due to these limitations, participants turn to human assistants. \rev{Human assistants} can interpret the context of tasks and goals more dynamically, \rev{providing} conversational guidance and adapting to users' actions to facilitate goal completion.

\paragraph{Example 1: Inadequacy in Identifying Central Puzzle Piece}

P14 used \sbma{} to differentiate between various puzzles by describing the images on the boxes like a scene of cats or bears. 
When tasked with identifying the centerpiece of a nine-piece puzzle using the box image, \sbma{} failed to provide the necessary detail for this precise task. 
Initially, the system correctly read text from the puzzle box indicating which piece belonged in the center. 
However, when P14 requested more specific guidance to locate the centerpiece, \sbma{} only noted that it was square -- a characteristic shared by multiple pieces. This response was too vague for successful puzzle assembly.

\begin{quote}
    \textit{``When \bma{} read me the text of the box on the back of the puzzle that said, you know, this particular piece should be in the center of the puzzle. As a follow-up question I asked, `Could I have more information about the piece in the center?' And it said, `This piece is a square piece,' but I mean, there were many different square pieces, so I could not tell from that.''} (P14)
\end{quote}
This example illustrates the system's limited ability to translate visual understanding into actionable guidance. While it could process the image and comprehend users' general goals, it could not provide the detailed information needed for successful goal achievement. 
As a result, P14 turned to a family member for assistance.

\begin{figure}[t!]
\centering
\includegraphics[width=0.6\columnwidth]{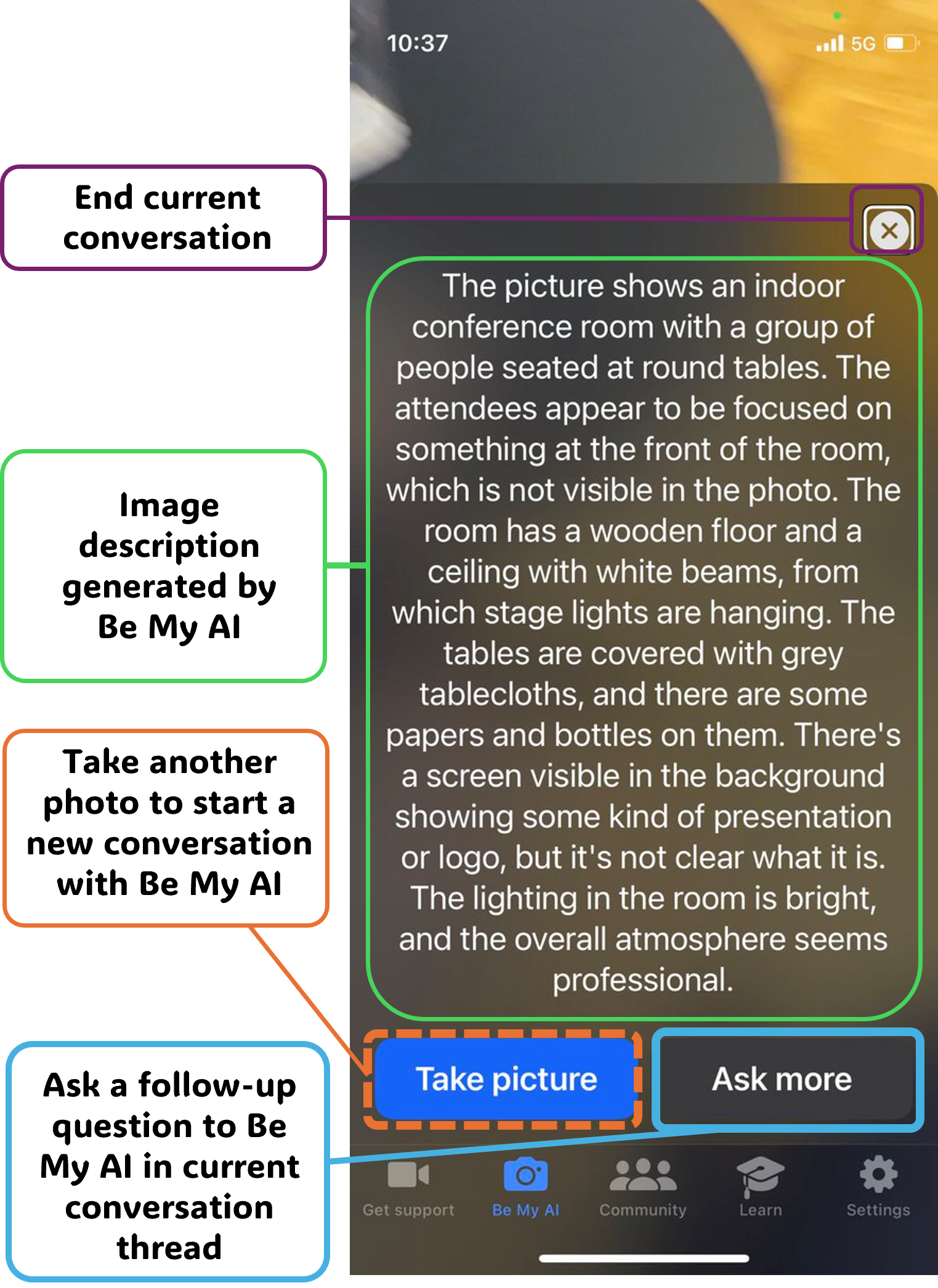}
\caption{\bma's description of a conference room, with the original image cropped. This example was drawn from X.}
\Description{Screenshot of the Be My AI interface. In the top is the description of a conference room. In the bottom left is the Take Picture button. In the bottom right is the Ask More button. The full description is as follows. The picture shows an indoor conference room with a group of people seated at round tables. The attendees appear to be focused on something at the front of the room, which is not visible in the photo. The room has a wooden floor and a ceiling with white beams, from which stage lights are hanging. The tables are covered with grey tablecloths, and there are some papers and bottles on them. There's a screen visible in the background showing some kind of presentation or logo, but it's not clear what it is. The lighting in the room is bright, and the overall atmosphere seems professional.}
\label{incomplete_info}
\end{figure}

\paragraph{Example 2: Lack of Instruction for Camera Adjustment}
\label{adjust_camera}

A common challenge for participants (P3, P5, P6, P8, P10-12, P14) was aligning the camera properly to capture clear views of their intended areas. 
\bma{} can identify image quality issues, notifying users when pictures were \textit{``cut off''} (P5), \textit{``blurry''} (P6), or incomplete (Figure~\ref{incomplete_info}).  
However, it is unable to provide further actionable guidance on adjusting the camera to improve image clarity. 
This limitation manifested in scenarios where \sbma{} recognized both the user's goal of capturing specific areas and the image deficiencies but could not suggest practical solutions. 
As P6 described, \textit{``It was very frustrating because it said the picture is blurry. Can you please put the label in the frame? But I didn't know how to put the label on the frame.''}

Eight participants (P1-3, P7, P8, P12-14) \rev{resolved this challenge by turning to human assistants, who can offer} adaptive support to achieve their goals, \rev{especially in} tasks that require continuous feedback. P13 shared an example where a human assistant not only understood her goal but also guided her in adjusting the camera and instantly reminded her to turn on the light to achieve her objective.

\begin{quote}
    \textit{``You know, having that ability to communicate and say, `Hey, this is what I'm looking for.' Or one time, I was looking for something and I had the lights off, and they're like, `You need to turn the lights on.' And I go, `Okay,' as opposed to, you know, if I tried using AI for that, it would just say `dark room.'''} (P13)
\end{quote}

This example highlights human assistants' ability to adapt their communication based on the situation and the user's implied needs, providing solutions that directly support achieving the user's goals.
\rev{However}, \bma{} struggles to provide practical assistance despite understanding basic requests.

\subsubsection{Real-Time Feedback}
\label{realtime_feedback}

We investigate how participants used \sbma{} to support navigation tasks, focusing on location awareness and orientation. 
Our analysis reveals the system's constraints in delivering real-time feedback and comprehensive navigational information from static images, and its inability to facilitate immediate interactions with surroundings.

\paragraph{Capabilities of \bma}

Participants (P2, P5, P10) employed \sbma{} to aid in localization and orientation while navigating to their destinations. 
For instance, P2 utilized it to identify gate numbers at the airport, P5 employed it to read signage directing toward the airport's transportation area, and P10 used it to recognize her surroundings when disoriented in her neighborhood.  
Despite these benefits, participants encountered challenges when using \sbma{} for navigation. The primary limitations stemmed from the limited camera view and practical mobility issues.

\paragraph{Example 1: Limited Navigational Information in Static Images}

Six participants (P2, P5, P6, P10, P11, P13) reported that \sbma's reliance on static images rather than real-time videos makes it difficult to capture comprehensive navigational information, such as obstacles and signage, in a single shot. 
As P6 described, users \textit{``have to stand there and keep taking pictures and taking pictures,''} verify the captured content, assess its utility for navigation, and adjust the angle for additional shots. This iterative process can be time-consuming, \textit{``It'll be too much of a task that's supposed to take maybe 10 minutes would probably take like 30.''}

P10 elaborated on the challenge and risk of simultaneously taking pictures and navigating, particularly when attempting to identify and navigate around obstacles. The uncertainty of capturing all potential hazards was a significant concern.

\begin{quote}
    \textit{``It's still hard to know how to capture, you know, all the obstacles. I think that's the issue. Like, how to know that I captured the right obstacle on my path? I mean, it depends [on] what I'm able to capture with the camera. You know, that's the tricky part for somebody without vision to capture the obstacle.''} (P10)
\end{quote}

Given these limitations, participants preferred human assistance through video-based interactions over \bma{} for real-time navigation. 
Video interactions not only benefit from human adaptability in adjusting guidance (Section~\ref{adjust_camera}) but also offer crucial real-time feedback unavailable in static images. 
Participants (P2, P6, P13) suggested that real-time video interpretation capabilities would significantly enhance the system by eliminating the need for repeated picture-taking.

\begin{quote}
    \textit{``If you could hold the camera and it could do it in real time, versus having to stop, take a [picture], then assess it... If that were the case, then you could move on to doing things like uploading videos and getting it to describe actual videos and things like that, versus just still images. That would be great. You know, then you could describe more.''} (P2)
\end{quote}
Such real-time video analysis would enable continuous camera movement and immediate feedback, streamlining the navigation process without pausing and reviewing individual images.

\paragraph{Example 2: Irreplaceable Role of Orientation \& Mobility Skills in Navigation}

Besides the value of real-time visual interpretations from external resources like human assistants, participants (P5, P10, P13, P14) emphasized the indispensable role of real-time feedback through their Orientation and Mobility (O\&M) skills for safe navigation.
P5 and P13 pointed out that even human assistance, though adaptive in guiding PVI away from navigational hazards, cannot substitute for essential O\&M skills required for tasks like street crossing.
Consequently, participants are more cautious about relying on emerging AI tools for navigation.
P14 reinforced this perspective, saying, \textit{``it's not a replacement for our mobility skills or just any skills in general. It can aid and augment the skills but it's not a replacement for them. It shouldn't be.''}

Participants (P5, P13) further delineated vital information provided by O\&M skills or discerned through O\&M tools like white canes or guide dogs, which current AI or human assistance cannot replace. 
First, immediate surroundings. \bma{} can identify obstacles \textit{``that are a little far away''} (P5) but may miss immediate surrounding hazards. 
Second, distance and proximity measurements of obstacles. Although \sbma{} can indicate the presence of obstacles, it lacks the capability to measure their distance.
Third, directional details. Essential navigational information, such as the direction of stairs (ascending or descending) or the presence of railings, are not always detectable through the system. 
With the O\&M skills and tools, users can instantly adapt their movements based on direct interaction with their environment. This level of responsiveness is currently unattainable with AI tools or human assistance.

\begin{quote}
    \textit{``[\bma] says, you know, `stairs in front,' and it's like, `Okay, that's great, but where are they? How far are they? Are they going up? Are they going down? Is there a railing?' which would be information that the dog or the cane could tell you. So, I would say use it as a tool along with, but definitely not by itself.''} (P13)
\end{quote}

In summary, neither AI systems nor human assistants can replace the essential, real-time feedback and adaptive capabilities offered by O\&M skills and tools. These elements are vital for ensuring safe navigation by allowing users to directly interact with their environment.

\section{Discussion}
\label{sec:discussion}

\begin{figure*}[t!]
\centering
\includegraphics[width=0.85\textwidth]{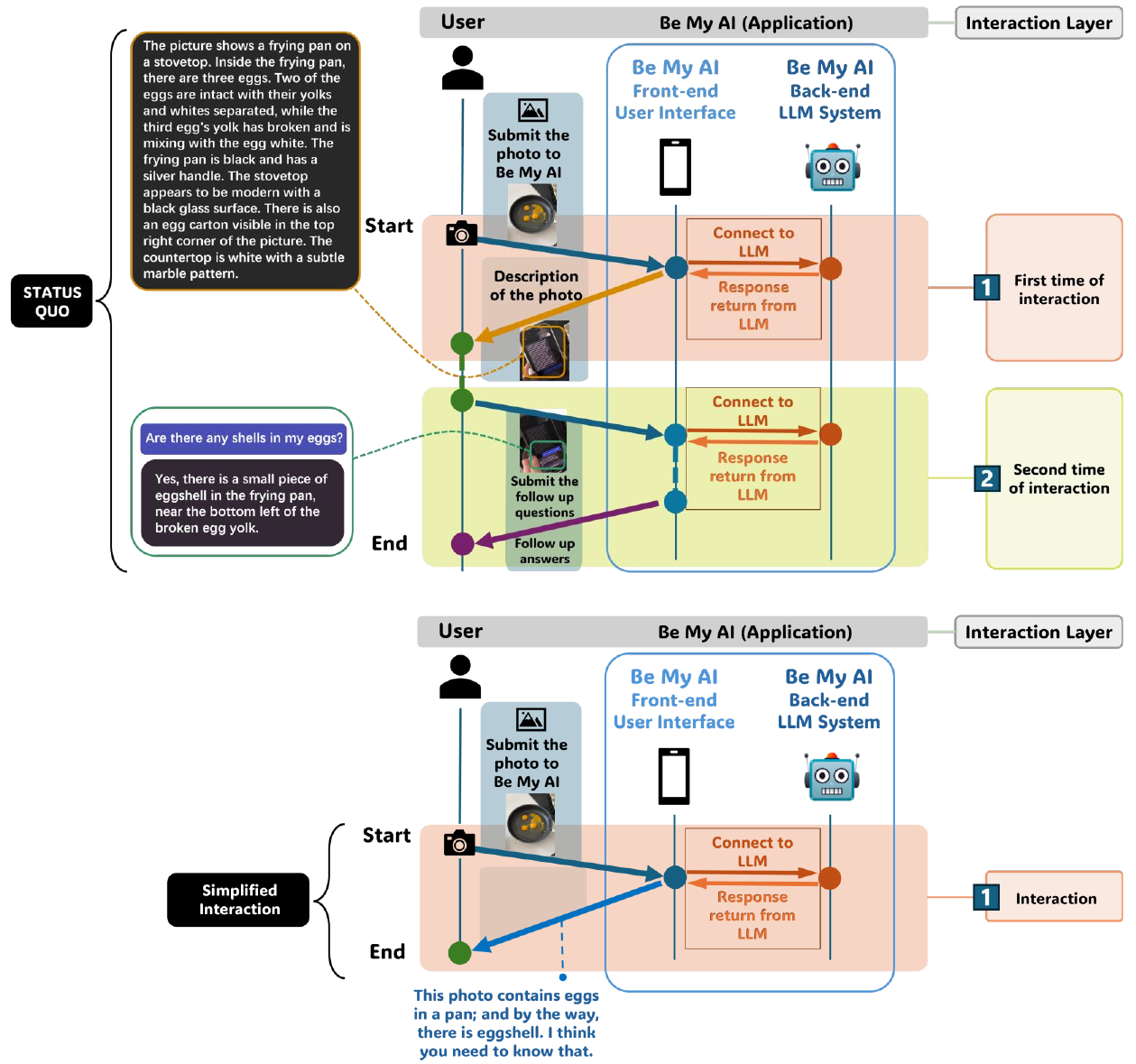}
\caption{The top shows the status quo of handoff between the user and \bma. The bottom illustrates our proposed simplified interaction.}
\Description{The top figure illustrates the current state of interactions between the user and Be My AI, where the user submits a photo and Be My AI returns a response in the first time of interaction. Subsequently, the user poses a follow-up question to Be My AI and Be My AI returns a more detailed description in the second time of interaction. The bottom figure is our proposed simplified interaction, where Be My AI learns from previous interactions and returns a detailed description during the first interaction, without user asking follow-up question.}
\label{fig.discussion_eggshell_user+app}
\end{figure*}

In this section, we examine the current state of handoff between users, \bma, and remote sighted assistants, and propose new paradigms to address the challenges identified in our findings. 
Next, we explore how multi-agent systems, both human-human and human-AI interactions, assist visually impaired users, and envision the transition toward AI-AI collaborations for tasks requiring specialized knowledge. Finally, we discuss the potential advantages of real-time video processing in the next generation of AI-powered VQA systems.

\subsection{Handoff Between Users, \bma, and Remote Sighted Assistants}
\label{handoff}

In this study, we illustrated the advantages of the latest LMM-based VQA system in (i) enhancing spatial awareness through detailed scene descriptions of objects' colors, sizes, shapes, and spatial orientations
(Section~\ref{physical_environments}), (ii) enriching users' understanding in social and stylistic contexts by detailing emotional states of people or animals, their body language, ambiance
(Section~\ref{social_stylistic}), and identity recognition (Section~\ref{identity}), 
and (iii) facilitating navigation by interpreting signages (Section~\ref{realtime_feedback}).

Informed by our findings, despite these various benefits, there are challenges that the system alone cannot overcome.  
\bma{} still requires human intervention, either from the blind user or the remote sighted assistant (RSA), to guide or validate its outputs.  
Users seek confirmation from RSAs or depend on their own spatial memory to overcome AI hallucinations, where \sbma{} inaccurately adds non-existent details to scenes. Users also rely on auditory cues and spatial memory to locate dropped objects and direct the system toward the intended search areas (Section~\ref{physical_environments}).
Moreover, users actively prompt the system to understand their specific objectives, such as checking for eggshells in a frying pan or adjusting appliance dials (Section~\ref{agentic_interaction}). 
There are also instances where users require assistance from RSAs when the system fails to provide adequate support to fulfill users' objectives, such as identifying the centerpiece of puzzles or adjusting the camera angle (Section~\ref{consistency}).
Human assistance or users' O\&M skills are necessary to receive real-time feedback for safe and smooth navigation (Section~\ref{realtime_feedback}).

Furthermore, our findings revealed that the system might produce inaccurate or controversial interpretations. Users express skepticism towards \sbma's subjective interpretations of animals' emotions and fashion suggestions (Section~\ref{social_stylistic}), and have encountered inaccuracies in \sbma's identification of people's gender and age (Section~\ref{identity}). These instances underline potential areas where human judgment is necessary to corroborate or correct the system's descriptions.

Next, we discuss the handoff~\cite{mulligan2020concept} between users, \bma, and RSAs to mitigate the aforementioned challenges.



\subsubsection{Status Quo of Interactions Between Users and \bma}

Through BMA's ``ask more'' function, users are able to request additional details about the images that were not covered in initial descriptions. 
This functionality facilitates a shift in interaction dynamics between users and \bma, even if the system may not accurately understand or answer users' questions in the first attempt. 
In these interactions, users are not merely passive recipients of AI-generated outputs, they actively guide the AI tool with specific prompts to better align AI's responses with their objectives.

Our findings reported one instance where the AI tool fails to grasp the user's intent to check the presence for eggshells in the beginning (Figure~\ref{fig.discussion_eggshell_user+app} top). First, the user submits an image of eggs in the pan. Respond to the image, the system describes the quantity and object (``Inside the frying pan, there are three eggs'') and states of the yolks and whites (``whites separated'', ``yolk has broken'', ``mixing with the egg white''). Next, the user clarifies her inquiry by asking, ``are there any shells in my eggs?''
This prompts the system to understand the user's goal, reevaluate the image, subsequently confirming the presence (``Yes, there is a small piece of eggshell in the frying pan'') and location of an eggshell to help her remove it (``near the bottom left of the broken egg yolk'').

This interaction exemplifies the status quo of handoff, where the user and \bma{} engage in a back-and-forth dialogue to refine the descriptions based on the ``ask more'' function and the user's precise prompts. 
\rev{While this iterative process allows the system to eventually understand the users' intent without RSAs' intervention, it places cognitive burden on users who must carefully craft and iteratively refine their prompts. The cognitive load increases as users mentally track what information they've already received, analyze gaps between their needs and the system's responses, and develop increasingly specific queries. 
}

\rev{To reduce users' cognitive load, we propose enabling the system to adopt a mechanism that combines multi-source data input with long-term and short-term memory capabilities~\cite{zhong2024memorybank}. With explicit user consent, future versions of LMM-based VQA systems could integrate data from users' mobile devices (e.g., location information, time data) alongside historical interaction data within the system (e.g., contexts and follow-up questions) to recognize user preferences and common inquiries, and infer their needs, thereby generating responses more effectively in similar contexts. Long-term memory serves as a repository for capturing generalized user preferences, behavior patterns, and aggregated insights across multiple users. This long-term memory is particularly effective for improving system intelligence by identifying common user needs and optimizing general responses~\cite{priyadarshini2023human}. Meanwhile, short-term memory can focus on task-specific optimization within a single interaction session. It retains context from the immediate conversation, such as recent user inputs and system responses, to enhance relevance and coherence in real time. Short-term memory operates dynamically, clearing retained data once the session ends or the task is completed, thereby ensuring privacy and preventing unnecessary data retention.}

\rev{For example, when identifying unusual elements during cooking (e.g., eggshells in cooking eggs), \bma{} could utilize the user's immediate input while referencing short-term memory from the current session or recent similar interactions. Additionally, by leveraging long-term memory, the system can learn from the user's past questions and query patterns to better match their habits and preferences, i.e., user's typical needs. Furthermore, multi-source data input, such as time or location information, can assist the system in inferring the user's current context, for instance, recognizing that the user is preparing a specific meal at a particular time or place, which allows the system to provide more relevant and context-aware assistance. This approach enables \bma{} to proactively anticipate user intent and deliver targeted responses, reducing the need for multiple clarifying prompts (Figure~\ref{fig.discussion_eggshell_user+app} bottom).} 

\rev{Cognitive Load Theory (CLT) suggests that well-designed interactions can significantly reduce users' extraneous load while enhancing the effective management of germane load~\cite{sweller1988cognitive, chandler1991cognitive}. Following the principles of CLT, we recommend using the above design to enable LMM-based VQA systems to minimize unnecessary clarifying prompts, thereby reducing users' cognitive load. 

}

\begin{figure*}[h!]
\centering
\includegraphics[width=0.8\textwidth]{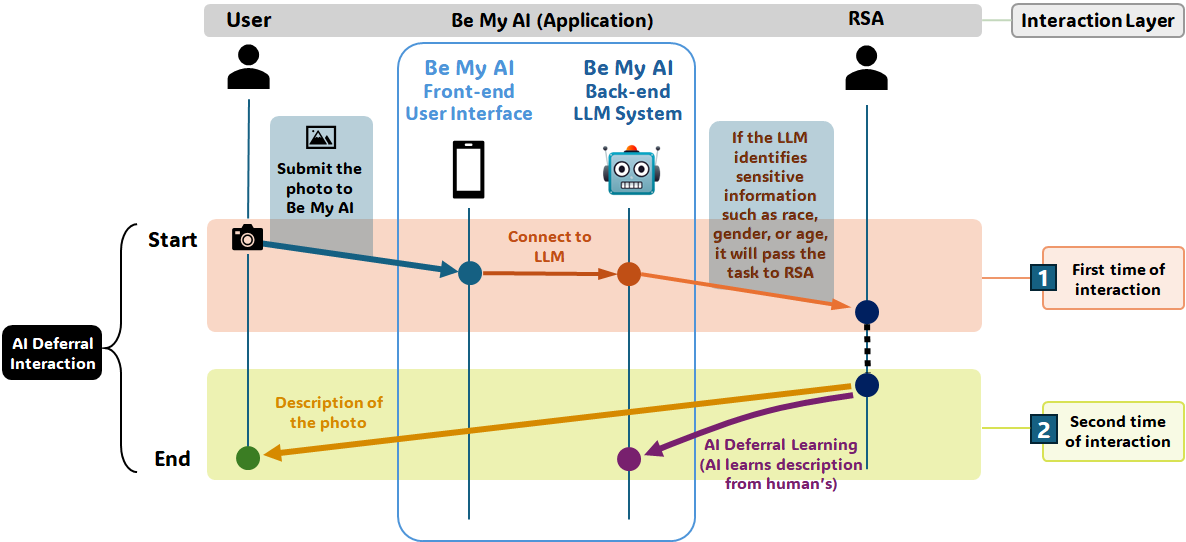}
\caption{Handoff between the user, \bma, and RSA for identity interpretations.}
\Description{The user submits a photo of people to Be My AI. Be My AI recognizes the requirement for identity interpretations and direct the photo to a remote sighted assistant. The remote sighted assistant returns a description of people's identities to Be My AI. Be My AI learns from the assistant's responses.}
\label{fig.discussion_eggshell_user+app+rsa}
\end{figure*}


\subsubsection{AI Deferral Learning for Identity Interpretations} 
\label{sec:deferral_learning}

Our findings elucidated \sbma's capabilities and limitations in interpreting identity attributes. Although the system can describe aspects like gender, age, appearance, and ethnicity, it may make errors due to its reliance on stereotypical indicators and its inability to interpret non-visible details (Section~\ref{identity}).
However, Stangl et al.'s work~\cite{stangl2020person} pointed out that PVI \rev{seek identity interpretations from AI assistants} across various contents, including \rev{browsing} social networking sites where our participants reported using \bma.

This reveals a tension between PVI's interests in knowing about \rev{identity} attributes and the AI's challenges in providing reliable information\rev{~\cite{hanley2021computer}}. The conflict arises because attributes such as age and gender are not purely perceptual and cannot be accurately identified by visual cues alone. \rev{However}, RSAs \rev{can draw on contextual clues, past interactions, and cultural knowledge to make more nuanced observations about these human traits.} These social strategies are not typically accessible to AI systems. 


To mitigate these issues, \rev{we propose adopting a deferral learning architecture~\cite{mozannar2020consistent, raghu2019algorithmic}, where an AI model learns when to defer decisions to humans and when to make decisions by itself. As detailed by Han et al.~\cite{han2024uncovering} and illustrated in Figure \ref{fig.discussion_eggshell_user+app+rsa}, this architecture creates a three-stage information flow:

\begin{itemize}
    \item \textbf{Stage 1}: It begins when users submit image-based queries to \bma. At this stage, the system uses a detection mechanism to identify sensitive contents, focusing particularly on those involving human physical traits. Current large-language models have already incorporated such mechanisms~\cite{perez2022red, bai2022training}; however, they still struggle to interpret human identity with consistent accuracy~\cite{hanley2021computer}.    

    \item \textbf{Stage 2}: Rather than declining sensitive requests outright, \bma{} redirects these queries to RSAs. This maintains the system's helpfulness while ensuring accurate responses. 
    \item \textbf{Stage 3}: RSAs provide descriptions by leveraging contextual understanding, such as analyzing the users' current environment and cultural background.
\end{itemize}

In contrast to prior work that addresses stereotypical identity interpretation through purely computational approaches~\cite{wang2019balanced,wang2020towards,ramaswamy2021fair}, our proposed AI deferral learning takes a hybrid human-AI approach that combines AI capabilities with human expertise. 
While previous AI-only solutions have made progress in reducing bias, they still struggle with identity interpretation~\cite{hanley2021computer}.
The challenges arise not only from technical issues but also from the ontological and epistemological limitations of social categories (e.g., the inherent instability of identity categories), as well as from social context and salience (e.g., describing a photograph of the Kennedy assassination merely as ``ten people, car'').
Our system leverages RSAs who have got more experience and probably more success in identifying people's identity through their human perception abilities and real-world experience. 
RSAs can interpret subtle contextual cues, understand cultural nuances, and adapt to diverse presentation styles that may challenge AI systems. 
Through the AI deferral learning architecture, AI assistants can learn continuously from human assistants' responses and improve its ability to handle similar situations. The three-way interactions between users, AI assistants, and RSAs generate rich contextual data that can enhance the AI system's identity detection mechanisms.}

\begin{figure*}[t!]
\centering
\includegraphics[width=0.8\textwidth]{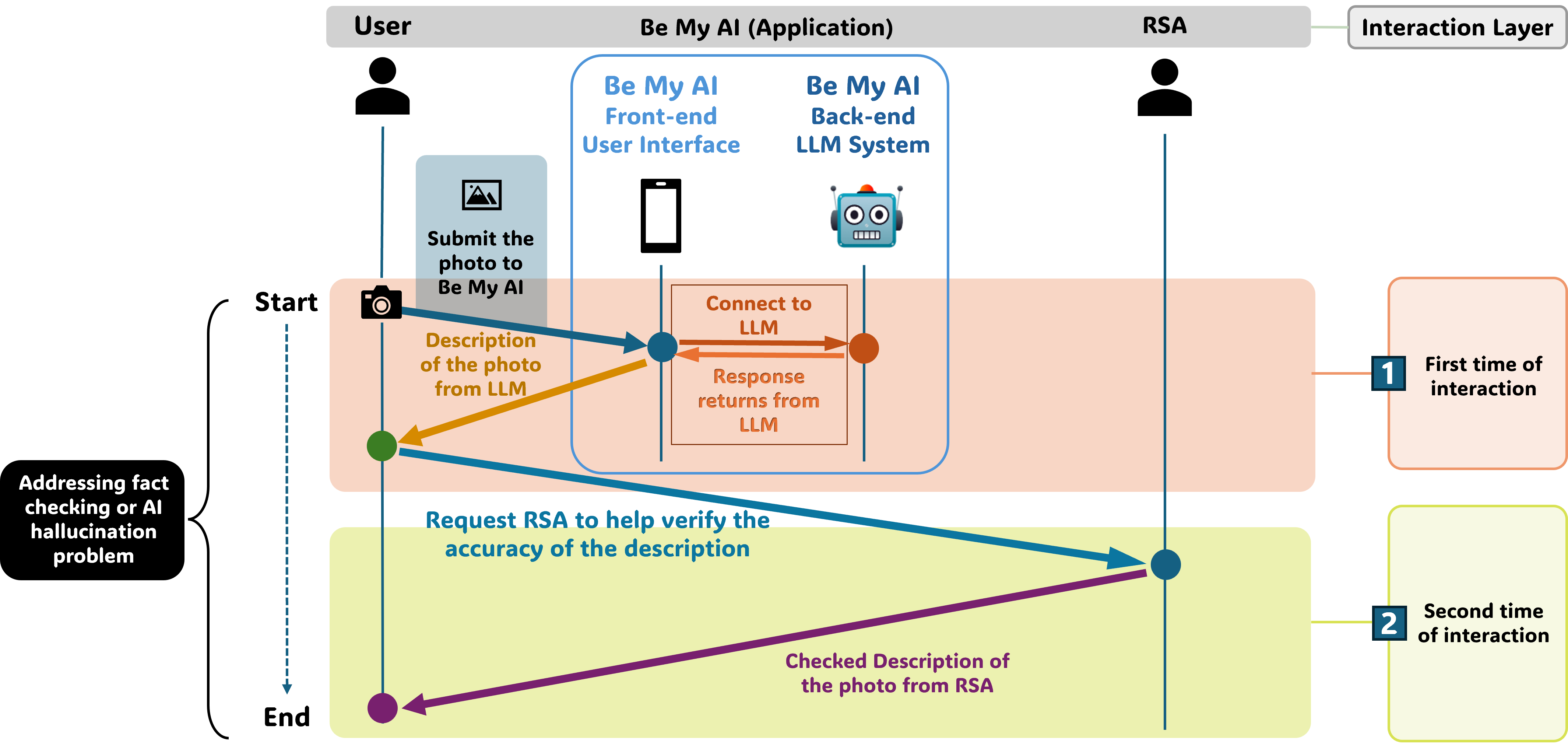}
\caption{Handoff between the user, \bma, and RSA for fact-checking.}
\Description{The user submit a photo to Be My AI, Be My AI returns a description with possible AI hallucination. Then, the user directs the photo to a remote sighted assistant, who returns a corrected description to the user.}
\label{fig.discussion_eggshell_user+app+rsa-check}
\end{figure*}

\subsubsection{Fact-Checking for AI Hallucination Problem}

Our findings highlighted that the AI-generated detailed descriptions helped users understand their physical surroundings. However, there were instances where AI systems hallucinated, i.e. incorrectly added non-existent details to the descriptions, which led to confusion (Section~\ref{physical_environments}). 
\rev{
In fact, hallucinations is a known problem for large language models upon which \bma{} is built~\cite{gonzalez2024investigating}. Current approaches to address this problem include Chain-of-thought (CoT) prompting~\cite{wei2022chain}, self-consistency~\cite{wang2022self}, and retrieval-augmented generation (RAG)~\cite{lewis2020retrieval}.

In CoT prompting~\cite{wei2022chain}, users ask an AI model to show its reasoning steps, like solving a math problem step by step rather than simply giving the final answer. It is similar to the ``think aloud'' protocol in HCI. 
Self-consistency~\cite{wang2022self} is an extension of CoT prompting. Instead of generating just one chain of thought, the model is asked to generate multiple different reasoning paths for the same task. Each reasoning path might arrive at a different answer. The model then takes a ``majority vote'' among these different answers to determine the final response.
In RAG~\cite{lewis2020retrieval}, AI models are provided with relevant information retrieved from a vector storage as ``context'' to reduce factual errors in their responses.
}

\rev{
In light of these techniques, users adopt various strategies that mirror CoT prompting, self-consistency, and RAG. We outline some potential strategies below.
\begin{itemize}
    \item \textbf{Part-Whole Prompting}: This strategy parallels the Chain of Thought (CoT) prompting. A user sends an image to \bma{} and requests an initial overall description, followed by a systematic breakdown that justifies this description. For example, users might first ask for a description of the image as a ``whole'', then request to divide the image into smaller ``parts'', like a $3 \times 3$ grid, and describe each grid individually. If the descriptions of the individual parts align coherently, it increases the likelihood that the overall description is accurate. This approach would require processing more information; however, it will provide users with greater confidence in the AI's response, as it enables them to verify consistency between the whole and its constituent parts.  

    \item \textbf{Prompting from Multiple Perspectives}: This strategy resembles the self-consistency technique. A user sends an image to \bma{} and requests multiple descriptions from different perspectives. For example, users might ask for one description that focuses on the background and another that emphasizes foreground objects. Users can also request descriptions from the viewpoint of objects within the image (e.g., ``How would a person sitting on a chair see this scene?'' and ``How would a person sitting on the floor see this scene?''). While gathering descriptions from multiple perspectives may increase the likelihood of hallucination, it can also help identify common elements that appear consistently across different viewpoints, potentially indicating true features of the image.    
    \item \textbf{Prompting with Human Knowledge}: This strategy resembles the RAG approach. A user sends an image to \bma{}, provides their current understanding of the image and its context, and requests a description that complements their knowledge. For example, in Figure~\ref{eggshells}, users can specify that someone took the picture in a kitchen environment and that it should show a frying pan containing eggs. Users possess this knowledge through their familiarity with physical environments, self-exploration, spatial memory, and touch~\cite{gonzalez2024investigating}. The user-provided knowledge will help the AI model ground its response in an accurate context~\cite{liu2024coquest}.     
    \item \textbf{Pairing with Remote Human Assistants}: While the previous three strategies rely on multiple prompting and response aggregation to identify facts, this approach leverages the traditional remote sighted assistance framework.
    This strategy (shown in Figure~\ref{fig.discussion_eggshell_user+app+rsa-check}) differs from the deferral learning framework (Section~\ref{sec:deferral_learning}) in that users forward the AI responses to human assistants, rather than the AI assistant deferring to humans for the response.
    In this strategy, a user first sends an image to \bma{} to receive a description. When users suspect inaccuracies through triangulation~\cite{gonzalez2024investigating}, such as descriptions that conflict with their spatial memory or common sense (e.g., implausible objects like a palm tree in a cold region), they can request a RSA to fact-check the description. The RSA then verifies the description and sends corrected information back to the user. This verification process is likely easier and faster for a RSA than composing a description from scratch, as the RSA's work involves checking rather than creating content.
\end{itemize}
}

\rev{
In summary, AI hallucination presents both challenges and opportunities. By addressing these issues, future work will strengthen the way users, AI models, and human assistants interact with each other.
}

\subsection{Towards Multi-Agent Systems for Assisting Visually Impaired Users}

This section examines the transition from human-human interactions to human-AI and AI-AI systems in supporting PVI. We explore how these multi-agent systems, which involve the collaborative efforts of multiple agents (AI or human), are designed to adaptively meet the diverse needs of PVI.

Lee et al.~\cite{lee2020emerging} identified four contexts in which a professional human-assisted VQA system (Aira) offer support to PVI. The type of information required by PVI is incremental in these contexts. 
First, \textit{scene description} and \textit{object identification} acquire information about ``what is it.''
Second, \textit{navigation} requires description about PVI's surroundings and obstacles (``what is it'') and directional information (``where is it'' and ``how to get to the destination'').
Third, \textit{task performance} like putting on lipstick, cooking, and teaching a class. This context requires description (``what is it'') and domain knowledge on ``how to do it.'' 
Forth, \textit{social engagement} like helping PVI in public spaces or interacting with other people. This needs description (``what is it''), directional information to navigate in social space, and discreet communication (PVI prefer not to disclose their use of VQA systems).

Our study reported how participants used \bma{} for tasks like matching outfits and assessing makeup, fitting under the category of \textit{task performance}. Some participants raised concerns about the accuracy of \sbma's interpretations and suggestions, indicating their preference for human subjectivity in this context. 
Contrasting with this, Lee et al.'s work~\cite{lee2020emerging} highlighted that remote sighted assistants (RSAs), even those professionals RSAs from Aira, sometimes lack the specialized information or domain knowledge required in task performance, thereby they need to collaborate with other RSAs to find solutions.

Furthering this investigation, Xie et al.~\cite{xie2023two} paired two RSAs to assist one visually impaired user in synchronous sessions, validating the need for RSAs to complement each other's description in task performance like aiding the user in applying makeup and matching outfits. They also explored the challenges in this human-human collaboration, revealing collaboration breakdowns between two opinionated RSAs. 
To address these issues, they proposed a collaboration modality in which one ``silent'' RSA supports the other RSA by researching but not directly communicating. This approach suggested that two RSAs in this multi-agent system should not deliver information simultaneously but have a clear division of labor, designating who takes the lead, to avoid overwhelming PVI with information.

Transitioning from human-human to human-AI collaboration, the handoff between the user, \bma{} and RSA (Section~\ref{handoff}) opens up new opportunities for multi-agent systems. 
Our proposed modality of human-AI collaboration integrates the scalable, on-demand capabilities of AI-based visual assistance with the contextual understanding and adaptability of RSAs. 
This multi-agent system involves the AI system recognizing its own limitations and seamlessly handing off tasks to a RSA when appropriate. This collaboration aligns with prior work~\cite{xie2023two}, where AI (\bma) and human (RSA) maintain a clear division of labor, minimizing cumbersome back-and-forth and reducing potential confusion for PVI.

Looking ahead, we envision the potential for AI-AI collaboration as part of the future multi-agent systems to assist PVI, especially for task performance. 
A domain-specific AI expert can be trained to handle more specialized tasks such as matching outfits, performing mathematical computations, or answering chemistry-related questions. \bma, as the core AI system, can provide general visual descriptions (``what is it'') and delegate more specialized tasks requiring domain knowledge to the domain-specific AI expert. 
This approach is in line with the human-AI collaboration (Section~\ref{handoff}) by ensuring effective handoffs when necessary. By leveraging AI agents with more specialized capabilities, this multi-agent system can better adapt to PVI's needs.

However, similar to concerns around human-human and human-AI interactions, these AI-AI collaborations must be carefully designed with clear protocols and handoff points for transitioning tasks between AI agents. It is important to make these transitions as seamless and transparent as possible to PVI, thereby avoiding any complexity or confusion.

\subsection{Towards Real-Time Video Processing in LMM-based VQA Systems}

One of the most significant advantages of \bma{} and other LMM-based assistive tools is their ability to provide contextually relevant and personalized assistance to users. By leveraging machine learning and natural language understanding, these systems can understand and respond to a wide range of user queries. This level of contextual awareness represents a significant advancement over pre-LMM-based assistive technologies, which often fail to adapt to the diverse needs and preferences of individual users.

However, our findings also identified several challenges and limitations associated with the reliance on static images by current LMM-based assistive tools. 
Participants in our study reported frustration with the need to take multiple pictures to capture the desired information, a process they found time-consuming and cognitively demanding (Section~\ref{realtime_feedback}). This iterative process hinders efficiency and also poses safety risks, as participants struggled with taking images while navigating around obstacles.

To mitigate these issues, integrating real-time video processing capabilities into future LMM-based VQA systems could offer significant benefits. 
Our findings suggest that the dynamic nature of video serves as a foundation for subsequent guidance (Section~\ref{realtime_feedback}), which is currently provided by human assistants through video-based remote sighted assistance.
Shifting to real-time video processing would allow LMM-based VQA systems to transition from identifying objects (answering ``what is it'') to offering practical advice (addressing ``how to do it''), such as how to adjust the camera angle or how to navigate to a destination.
By continuously analyzing the user's surroundings through real-time video feeds, these systems can dynamically interpret changes and provide immediate feedback, thus eliminating the need for static image captures. This capability would enhance the user experience by offering seamless navigation aid in real time.

The feasibility of real-time video processing is supported by existing technologies demonstrated in commercial products and research prototypes. For instance, systems that utilize sophisticated algorithms for real-time object segmentation in video streams~\cite{wang2021swiftnet} have shown significant potential in other domains. Building on these techniques for video analysis could significantly extend the capabilities of future LMM-based VQA systems.

Transitioning from static image analysis to real-time video processing can alleviate the burden of iteratively taking pictures and adjusting angles experienced by users. It can also enhance the utility and safety of LMM-based VQA systems, particularly during navigation. 
This progression, driven by ongoing advancements in machine learning and computer vision, is essential for the development of more adaptive and responsive assistive technologies that align with the dynamic nature of real-world environments.

\subsection{Limitations}

There are \rev{some} limitations of this work.
First, we completed the interviews prior to March 2024 and gathered \sbma's image descriptions from its official release until March 31, 2024. Thus, our analysis did not encompass versions of \bma{} released after March 2024. Given the frequent updates and evolution of the system, there may be feature changes or enhancements introduced post-March 2024 that were not considered in our study. Future research can investigate the capabilities and limitations of newer versions of \bma{} to provide updated insights.  
\rev{Second, while our focus on initial user experiences provided insights into early-stage interactions with \bma, this early-phase data collection resulted in a limited dataset. Future work can incorporate larger real-world datasets over extended periods of use and explore more diverse use cases to provide a more comprehensive understanding of the system's capabilities and limitations.}
\rev{Third}, the interview participants live inside the United States, and only image descriptions written in English were collected and analyzed. Thus, this study may not reflect perspectives outside the United States or from non-English speaking contexts. Future research can explore the usage and limitations of LMM-based VQA systems across broader and diverse cultural backgrounds. 
\rev{Fourth}, \bma{} cannot store conversation history. This limited functionality prevented users from sharing their ongoing interactions with the system. Thus, we were unable to fully understand users' prior experiences or analyze the data within broader contexts.

\section{Conclusion}

This study investigates the application of LMMs, particularly through \bma, to enhance accessibility for PVI. Our research explores both the capabilities and limitations of the system by interviewing 14 visually impaired users and analyzing image descriptions generated by it. 
We identify significant limitations in \bma's context-aware and intent-oriented capabilities, including AI hallucinations, subjective interpretations in social and stylistic contexts, inaccurate recognition of people's identities, and inconsistent support in understanding and acting on user intentions. 
These challenges often lead users to rely on human assistance or personal strategies to compensate. 
Informed by these findings, we propose approaches to enhance interaction between PVI, \bma, and remote sighted assistants, emphasizing streamlined interactions, more accurate identity recognition, and reduced AI errors. We also highlight the potential of multi-agent systems, fostering collaboration among humans and AI, and suggest exploring AI-AI cooperation for tasks requiring specialized knowledge.

\begin{acks} 
We thank Samantha Paradero for her assistance in data collection.  
We also thank the anonymous reviewers for their insightful comments. 
This research was supported by the US National Institutes of Health, and the National Library of Medicine (R01 LM013330). 
\end{acks} 

\bibliographystyle{ACM-Reference-Format}
\bibliography{bibliography}

\clearpage
\appendix
\onecolumn
\section{Appendix}
\label{appendix}


\begin{table}[h]
\caption{Codebook with Category, Example Tasks, and Codes}
\begin{tabular}{>{\color{black}}p{2.2cm}>{\color{black}}p{6cm}>{\color{black}}p{3.5cm}>{\color{black}}p{3.5cm}} 
\toprule
\textbf{Category}    & \textbf{Explanation}    & \textbf{Example Tasks}    & \textbf{Codes}        \\ \toprule
Physical Environment Understanding         & \bma{} enhances spatial awareness in both indoor and outdoor settings through detailed and structured visual information. Challenges include AI hallucinations that require verification through user's existing spatial knowledge or human assistance, and the need for users to complement the tool with their auditory perception for accurate object location. & Describe scenes like theaters, room layouts, neighborhood, and holiday decorations; Locate dropped objects like earbuds                                         & Environmental familiarity, Object location identification, Concerns in AI hallucinations, Human verification needs, Multi-modal perception integration              \\ \hline
Subjective Interpretations and Suggestions & \bma{} provides subjective interpretations in animal interaction and fashion contexts, such as animals' emotional states and fashion suggestion. Challenges includes lack of accuracy or context sensitivity, leading users to rely on personal judgment or sighted assistance for validation.                                                                     & Human-animal interactions like taking picuters of birds, dogs, horses; Fashion help like describing or matching colors and patterns of outfits, checking makeup & Subjective interpretation capability, Untrust in subjective interpretations, Agency in personal judgment                                                          \\ \hline
Identity Recognition                       & \bma{} enhances the perception of people by conveying identity attributes like age, gender, and ethnicity in image descriptions. Challenges in accuracy underscore the need for sensitivity and contextual awareness in identity depiction.                                                                                                                        & Human social interactions like describing online photos, checking photos before posting online                                                                  & Identity attribute recognition, User concerns in identity recognition                                                                                               \\ \hline
Goal Understanding Dialogue       & \bma's ``ask more'' function enables users to inquire specific information, supporting independent information gathering. However, \bma{} often requires explicit user guidance to focus on relevant details, as it struggles to infer user intentions from initial queries.                                                                         & Cooking and dining like checking eggshells; Helping with household appliances like rotating dials on washers and thermostats                                    & Query refinement capability, Intention inference limitations, User prompting for goal understanding                                                                    \\ \hline
Goal Achievement Support                   & \bma{} understands users' goals but faces challenges in proactively suggesting further actions to align with the goals, requiring human assistance to maintain focus and achieve goals.                                                                                                                                                                            & Identifying central puzzle piece, Adjusting camera angle                                                                                                        & Proactive guidance limitations, Human assistance for continuous and actionable guidance                                                                             \\ \hline
Real-time Navigation Assistance            & \bma{} assists with localization and orientation tasks through static image interpretations, but shows limitations in providing real-time and comprehensive navigational feedback. Users often need to complement its capabilities with human assistance and O\&M skills for safe navigation.                                                                      & Navigating users in airports like reading signage and gate numbers; Finding restaurants                                                                         & Localization and orientation capability, Static image limitations, Navigational risks, Human assistance for real-time interactions, Essential role of O\&M skills  \\ \bottomrule
\end{tabular}
\Description[Codebook]{The table is a structured summary categorizing themes related to the functionality and challenges of Be My AI. It includes columns titled Category, Explanation, Example Tasks, and Codes. The themes covered are: Physical Environment Understanding, Subjective Interpretations and Suggestions, Identity Recognition, Interactive Query Response, Goal Achievement Support, and Real-time Navigation Assistance. Each theme is further broken down into specific codes and example tasks, illustrating how Be My AI operates and interacts in various scenarios.}
\label{codebook}
\end{table}

\end{document}